\documentclass[twocolumn]{jpsj3}
\usepackage{txfonts}
\usepackage{color}

\title{Spin Dynamics at Very Low Temperature in Spin Ice Dy$_2$Ti$_2$O$_7$}

\author{
\name{Kazuyuki \surname{Matsuhira}}\thanks{E-mail: matuhira@elcs.kyutech.ac.jp}, \name{Carley \surname{Paulsen}}$^{1}$, \name{Elsa \surname{Lhotel}}$^{1}$, \name{Chihiro \surname{Sekine}}$^{2}$, \name{Zenji \surname{Hiroi}}$^{3}$, and \name{Seishi \surname{Takagi}}
} 

\inst{
\address{Kyushu Institute of Technology, Kitakyushu 804-8550, Japan}\\
\address{$^1$ Institute N\'{e}el C.N.R.S - Universite Joseph Fourier, BP 166, 38042, Grenoble, France}\\
\address{$^2$ Muroran Institute of Technology, Muroran, Hokkaido 050-8585, Japan}\\
\address{$^3$ Institute for Solid State Physics, University of Tokyo, Kashiwa 277-8581, Japan}\\
}

\abst{
We have performed AC susceptibility and DC  magnetic relaxation measurements on the spin ice system Dy$_2$Ti$_2$O$_7$ down to 0.08 K.
The relaxation time of the magnetization has been estimated below 2 K down to 0.08 K.
The spin dynamics of Dy$_2$Ti$_2$O$_7$ is well described by using two relaxation times ($\tau_{\rm S}$ (short time)  and  $\tau_{\rm L}$ (long time)).
Both $\tau_{\rm S}$ and $\tau_{\rm L}$ increase on cooling.
Assuming the Arrhenius law in the temperature range 0.5-1 K, we obtained an energy barrier of 9 K.
Below 0.5 K, both $\tau_{\rm S}$ and $\tau_{\rm L}$ show a clear deviation from the thermal activated dynamics toward temperature independent relaxation, suggesting a quantum dynamics.
}

\kword{Spin ice, pyrochlore, dynamics, magnetization, AC magnetic susceptibility, relaxation curve}

\begin{document}
\maketitle

The pyrochlore oxide Dy$_2$Ti$_2$O$_7$ is considered to be a typical example of a spin ice compound.~\cite{SpinIceRev,DyTiO-S,DyTiO-Neutron}
In this compound, Dy$^{3+}$ ions occupy the vertices of corner-shared tetrahedra.
The magnetic moment of Dy$^{3+}$ has a strong single-site anisotropy along the local $\langle 111\rangle$ axes, which is caused by crystalline electric field effects (CEF) and gives rise to a Kramers ground state doublet.
The strong dipolar interaction between Dy$^{3+}$ ions associated with the weak antiferromagnetic superexchange interaction results in an effective ferromagnetic nearest neighbor interaction $J_{\rm eff}$ estimated to be 1.1 K.~\cite{SpinIceRev}
The ferromagnetic interaction in combination with the strong single-site anisotropy stabilizes the local spin arrangement in such a way that the ground state configuration has  two spins pointing out of the tetrahedron and two spins pointing into the tetrahedron, (the so-called `two-in two-out' state or 'ice rules').~\cite{SpinIce} 
For every tetrahedron, there are six possible combinations of spins under the two-in two-out rule reflecting the global cubic symmetry. 
In fact, Dy$_2$Ti$_2$O$_7$ is found to show the residual ground state entropy of 1.68 J/(K mole), which is numerically in good agreement with the Pauling's entropy for water ice.~\cite{DyTiO-S,Ice} 
The spin ice state is attained by the development of a short-ranged spin correlations.

Recent theoretical studies reveal that the excitations from the ground state can be described by the creation of magnetic monopoles.~\cite{monopole}
Violating the ice rules by making a spin flip on the ground state configuration leads to a pair of point-like defects in the tetrahedra that contain the spins.
In fact, various experimental results are explained by the excitation of magnetic monopoles.~\cite{monopole-Morris, monopole-Fennell, monopole-Kadowaki, monopole-Bramwell, monopole-dynamics}
 
The spin dynamics of Dy$_2$Ti$_2$O$_7$ revealed by AC susceptibility $\chi_{\rm AC}$ measurements down to 0.6 K show a quite unique behavior, which can be roughly described by three regimes.~\cite{DyTiO-AC,DyTiO-AC2,DyTiO-AC3}
Above $\sim$ 10 K, the temperature dependence of the relaxation time $\tau(T)$ is effectively explained on the basis of the Arrhenius law $\tau(T) = \tau_{th} \exp(E_{\rm B}/T)$, with an energy barrier $E_{\rm B}$ = 220 K.~\cite{DyTiO-AC} 
In the intermediate regime, in the temperature range of 2-10 K, $\tau(T)$ is almost constant, and  below 2 K, $ \tau(T)$ increases again and reaches  $\sim 1$ s at 0.75 K.~\cite{DyTiO-AC3}
A difference between the zero field cooled (ZFC) and field cooled (FC)  DC magnetization is observed below 0.7 K.~\cite{DyTiO-AC3,DyTiO-DC}
These dynamics of spin ice may be understood by two origins.
At high temperature above $\sim$ 10 K, the dynamics is due to single ion process and mixing with excited states.~\cite{HotSpinIce}
$E_{\rm B}$ corresponds to the energy of the excited CEF levels which results in  the Ising-like anisotropy.
At low temperature below $\sim$ 10 K, the dynamics comes from the creation or annihilation of magnetic monopoles, and their diffusion.
Within the nearest neighbor spin ice model, the energy cost of a single spin flip is $4J_{\rm eff}$.
However, the experimental data of $\tau(T)$ can be well fit by an Arrhenius law with an energy barrier of $E_{\rm p} = 2 J_{\rm eff} \sim 2.2$ K over the restricted temperature range between 2.5 and 5 K.\cite{DyTiO-AC3, monopole-dynamics}
In a similar way, the low temperature increase of $\tau(T)$ can also be fitted by an Arrhenius law but with $E_{\rm p} = 6 J_{\rm eff} \sim 6.6$ K in the temperature range 2 K down to 0.75 K.
This behavior has been explained by taking into account the long range Coulomb interaction between monopoles.
In addition, a detailed analysis of diffusive motion of monopoles using classical Monte Calro simulation can reproduce the $\tau(T)$ dependence down to 0.75 K.\cite{monopole-dynamics,plateau}

Below 0.7 K, the spin dynamics were probed by magnetocaloric measurements,~\cite{DyTiO-MagCalo} but the  large applied fields make the comparison with $\chi_{\rm AC}$ data quite difficult.
Short time relaxation of the magnetization has been measured down to 0.36 K and could be accounted for within a chemical kinetic model.~\cite{DyTiO-Giblin}
To get a deeper insight in the spin dynamics of Dy$_2$Ti$_2$O$_7$, we have performed $\chi_{\rm AC}$ measurements and DC relaxation measurements down to 0.08 K.
In this letter,  we show that the magnetic relaxation curves as well as the $\chi_{\rm AC}$ data is well described by using two relaxation times down to very low temperature.
Furthermore, we show that $\tau(T)$ flattens at very low temperature, thus deviating from the suggested Coulomb model.~\cite{monopole-dynamics}

Single crystals of Dy$_2$Ti$_2$O$_7$ were prepared by the floating-zone method, as reported previously.~\cite{KagomeIce} 
$\chi_{\rm AC}$ and DC magnetic relaxation along the [111] direction were measured using a SQUID magnetometer developed at the Institut N\'{e}el, CNRS, Grenoble.
Two samples were used to ensure the experimental results; the size of the samples (\#1 and \#2) were $0.7 \times 2.2 \times 2.2$ ${\rm mm}^3$ and $0.9 \times 1.9 \times 3.8$ ${\rm mm}^3$, respectively.
No meaningful sample dependence is observed.
In this letter, the data of $\chi_{\rm AC}$ for sample \#1 and DC magnetic relaxation measurements for sample \#2 are shown; the data of DC magnetization at 0.08 K is shown for sample \#1.
The $\chi_{\rm AC}$ measurements were performed using AC magnetic fields $<1.4$ Oe.

In order to minimize the demagnetizing field corrections, the [111] direction was oriented along the sample plane.
The demagnetization correction for the data of $\chi_{\rm AC}$ is performed using the factor $N = 0.197\times 4 \pi$ (CGS units) that was calculated with the analytical form for a rectangular prism.~\cite{DyTiO-AC,Demag}
The relaxation of the magnetization was made using the following protocol; (i) a weak DC magnetic field of 10 or 5 Oe is applied. (ii) the sample is warmed up to 0.90 K for a wait period of 30 s. (iii)  the heater power was cut, and the sample was subsequently rapidly cooled down, dropping below 0.4 K in less than 10 s and settling down to below 0.1 K after 600 s (iv) the heater power was restored and regulated and stabilized at the target temperature for an additional 600 s. (v) Finally the applied field is cut (in a time $t < 0.1$ s) and the relaxation of the DC magnetization $M(t)$ is measured.

\begin{figure}[htbp]
\begin{center}
\includegraphics[width=7cm]{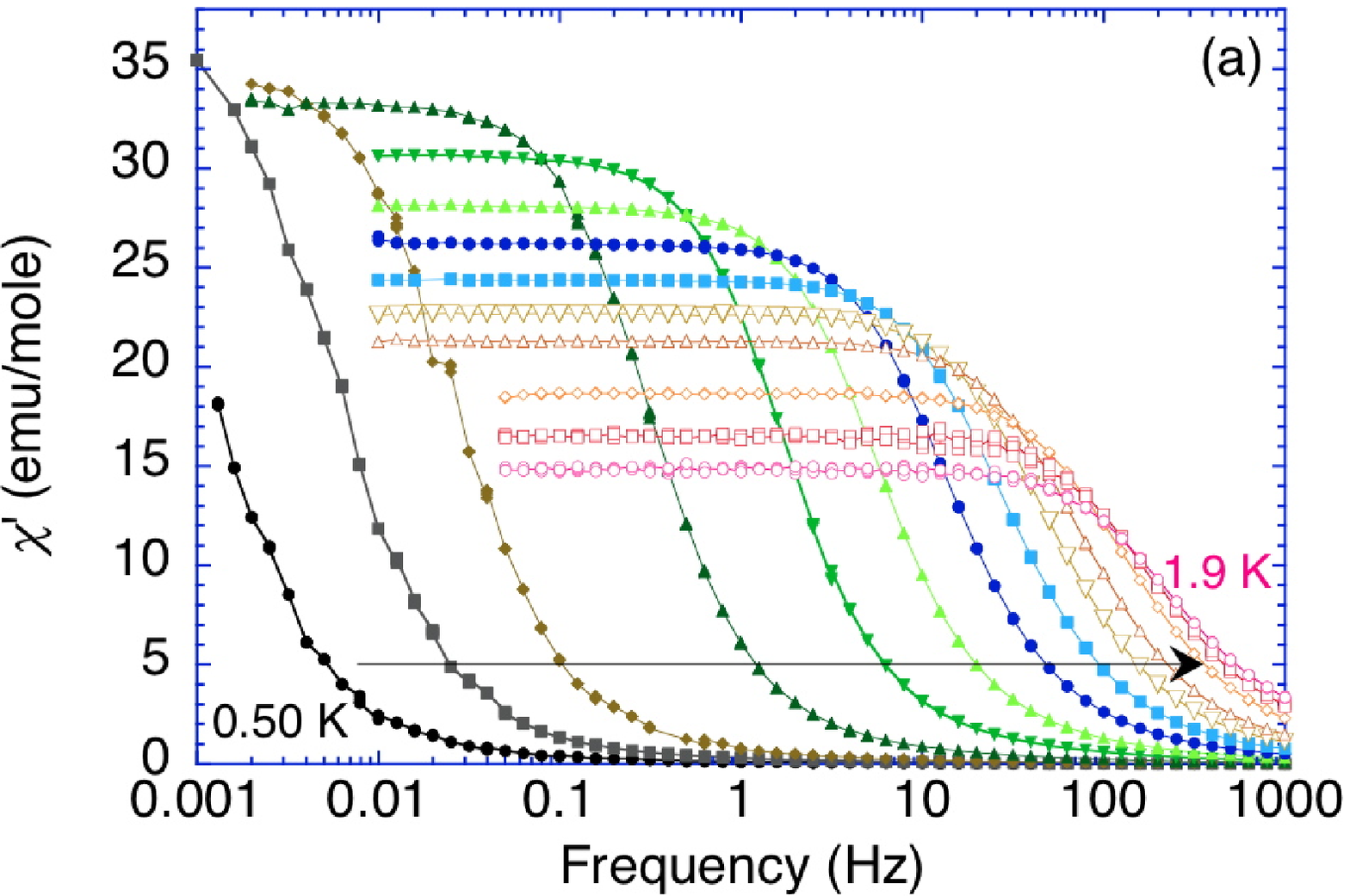}\\
\includegraphics[width=7cm]{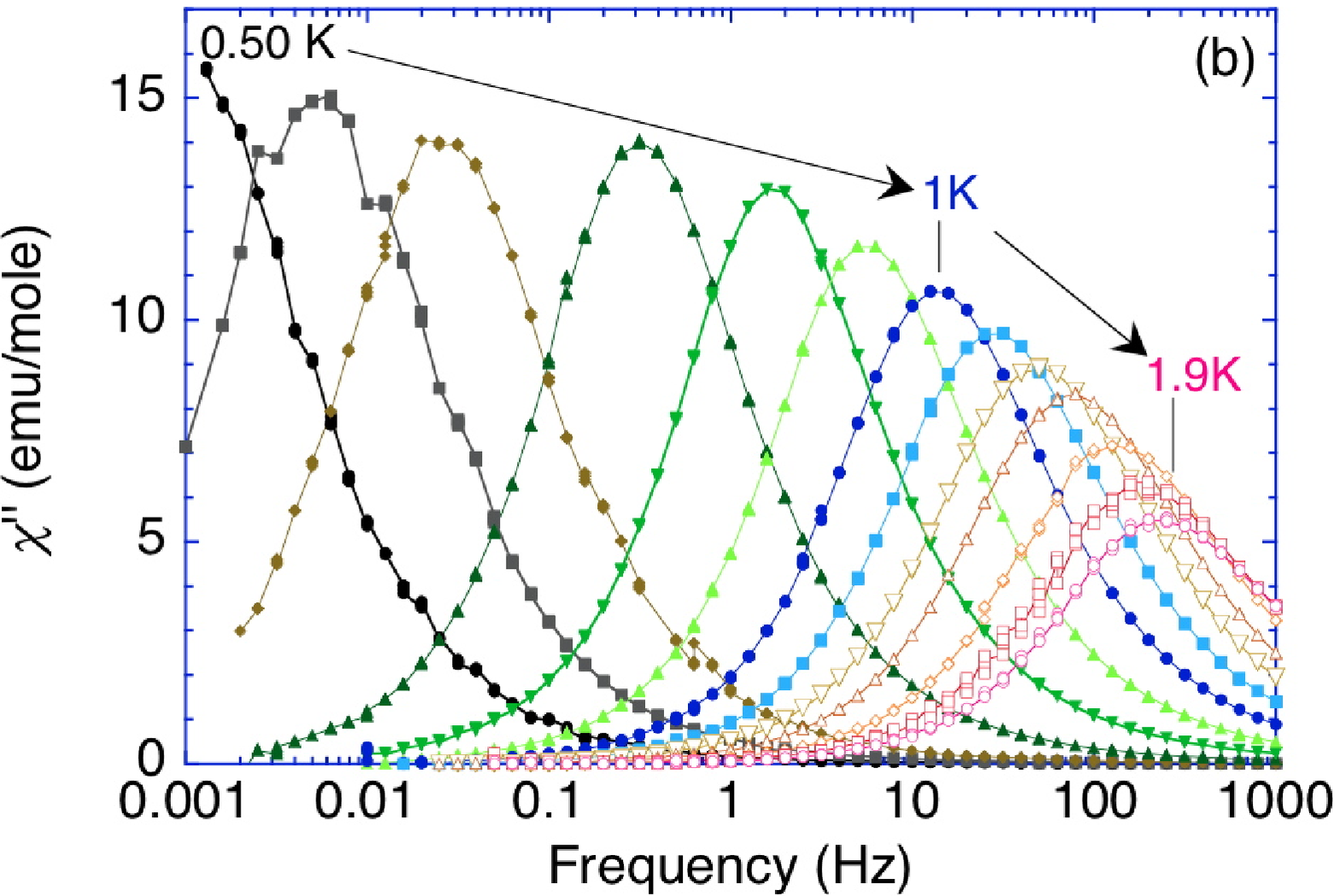}
\end{center}
\caption{
(a)(Color online)  Frequency dependence of $\chi^{\prime}(f)$ of Dy$_2$Ti$_2$O$_7$ at 0.50, 0.55, 0.60, 0.70, 0.80 0.90. 1.0, 1.1, 1.2, 1.3, 1.5, 1.7, and 1.9 K. (b)(Color online)  Frequency dependence of $\chi^{\prime\prime}(f)$ of Dy$_2$Ti$_2$O$_7$ at 0.50, 0.55, 0.60, 0.70, 0.80 0.90. 1.0, 1.1, 1.2, 1.3, 1.5, 1.7, and 1.9 K.
}
\label{f1}
\end{figure}

Clearly the most important parameter for the study of the dynamic behavior is the relaxation time $\tau(T)$, which can be deduced from $\chi_{\rm AC}$ as well as the relaxation of the magnetization.
In the case of Debye relaxation with a single dispersion, the imaginary part of the AC susceptibility $\chi^{\prime\prime}$ is Lorentzian, which is a symmetric function of log$f$, and a maximum occurs when the measurement time $1/2\pi f$ is equal to the relaxation time $\tau$.
In addition, the magnetization is expected to decay as an exponential function of time, with a characteristic time equal to $\tau$.
The frequency dependence of $\chi_{\rm AC}$ in the temperature range from 0.5 to 1.9 K is shown in Fig. \ref{f1}.
Our data are in qualitative agreement with a previous study made on a polycrystalline sample between 0.8 and 1.8 K.~\cite{DyTiO-AC3, Previous-data}
The position of the peaks in  $\chi^{\prime\prime}(f)$ shift to  lower frequency with decreasing temperature, implying  that $\tau$ increases on cooling as expected for a thermally activated process. 
The shape of the observed $\chi^{\prime\prime}(f)$ curve is very close to a Lorentzian symmetric function so that the main part of the dissipation can be attributed to a single relaxation time.
However the curves do show a small but clearly discernible shoulder on their high frequency sides, indicating the existence of a second relaxation time, or perhaps a distribution of relaxation times. 
Note that a similar feature in $\chi_{\rm AC}$ was observed for the high temperature data ($T > 14$ K).
The data have been  analyzed using the  Davidson-Cole formulation with a cutoff time, as previously pointed out.~\cite{DyTiO-AC,DavidsonCole}

\begin{figure}[htbp]
\begin{center}
\includegraphics[width=7cm]{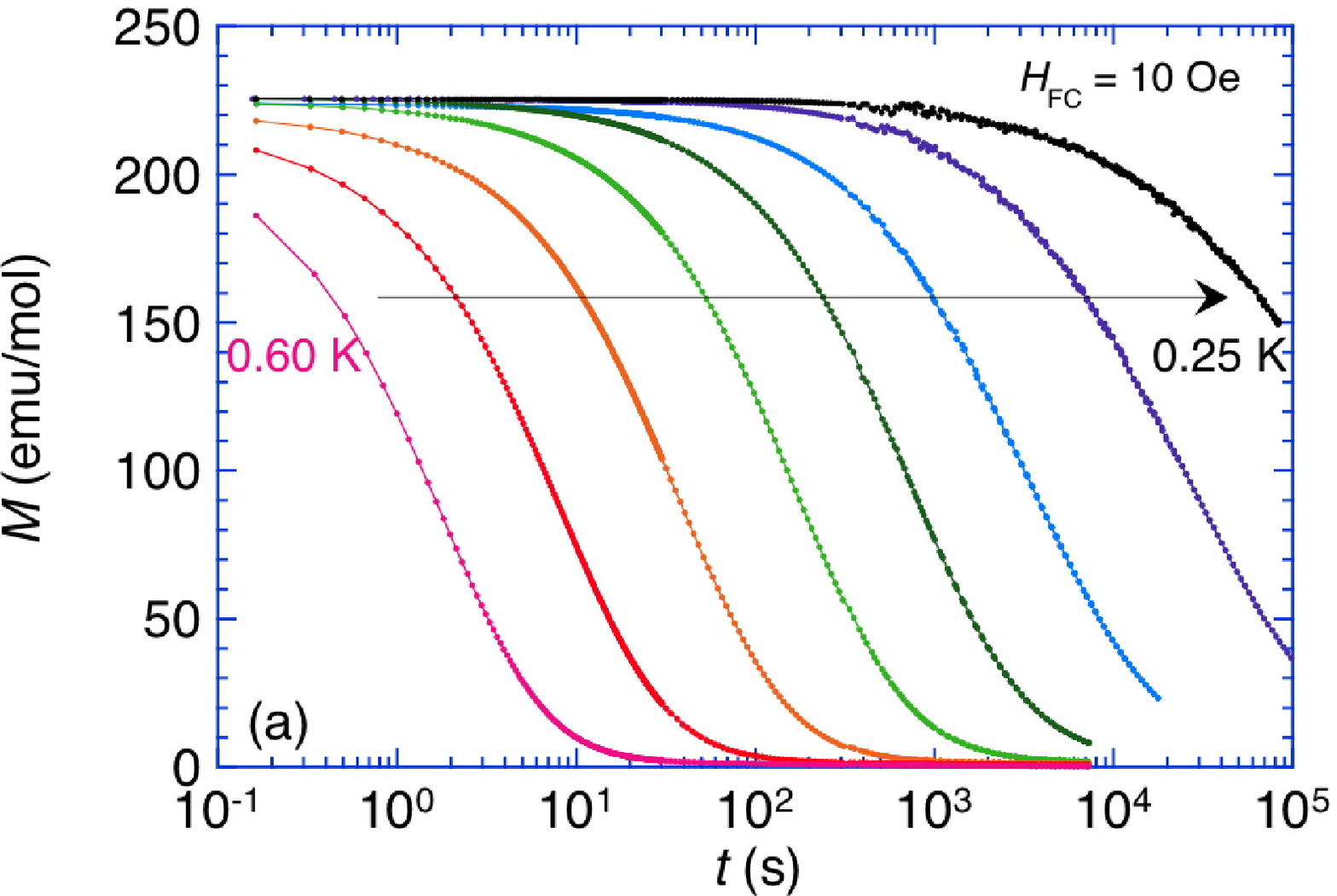}\\
\includegraphics[width=7cm]{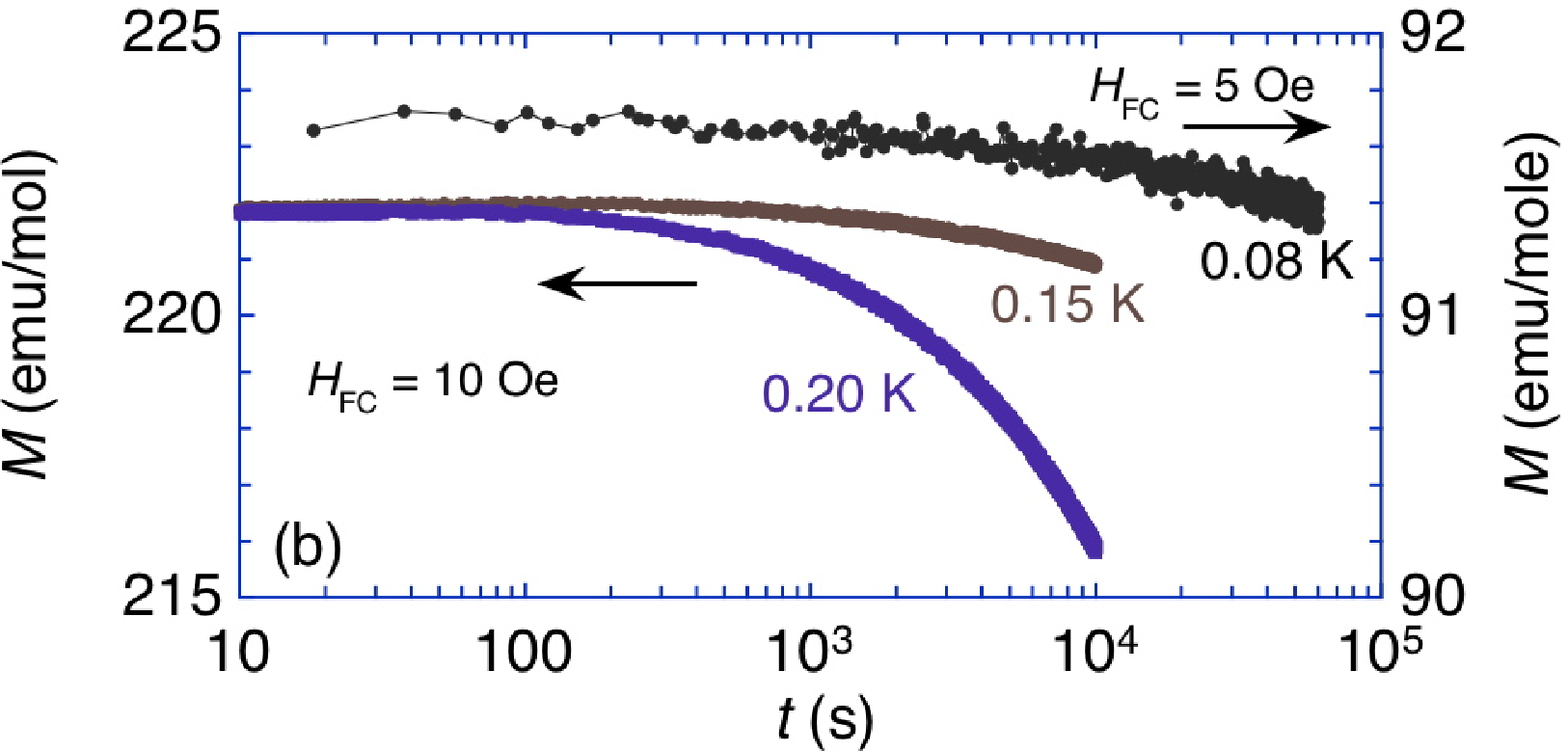}
\end{center}
\caption{
(Color online) Relaxation of magnetization due to field quench after field cooling (FC) at (a)  0.25, 0.30, 0.35, 0.40, 0.45, 0.50, 0.55, and 0.60 K and (b) 0.20, 0.15, and 0.08 K. 
The applied magnetic fields in FC $H_{\rm FC}$ for the data at 0.15-0.60 and  0.08 K are 10 and 5 Oe, respectively.
}
\label{f2}
\end{figure}

These observations are in agreement with the analysis of $M(t)$ curves shown in Fig. \ref{f2} below 0.60 K.
They were obtained (as described above) by the FC condition and then quenching the field at the measurement temperature.
As observed in $\chi_{\rm AC}$, the relaxation times increase at lower temperature.
Note that, at 0.08 K, the magnetization is reduced by only 0.35 \% after 16.7 h (See Fig. \ref{f2}(b)), showing that slow relaxation still occurs at this temperature.

\begin{figure}[htbp]
\begin{center}
\includegraphics[width=7cm]{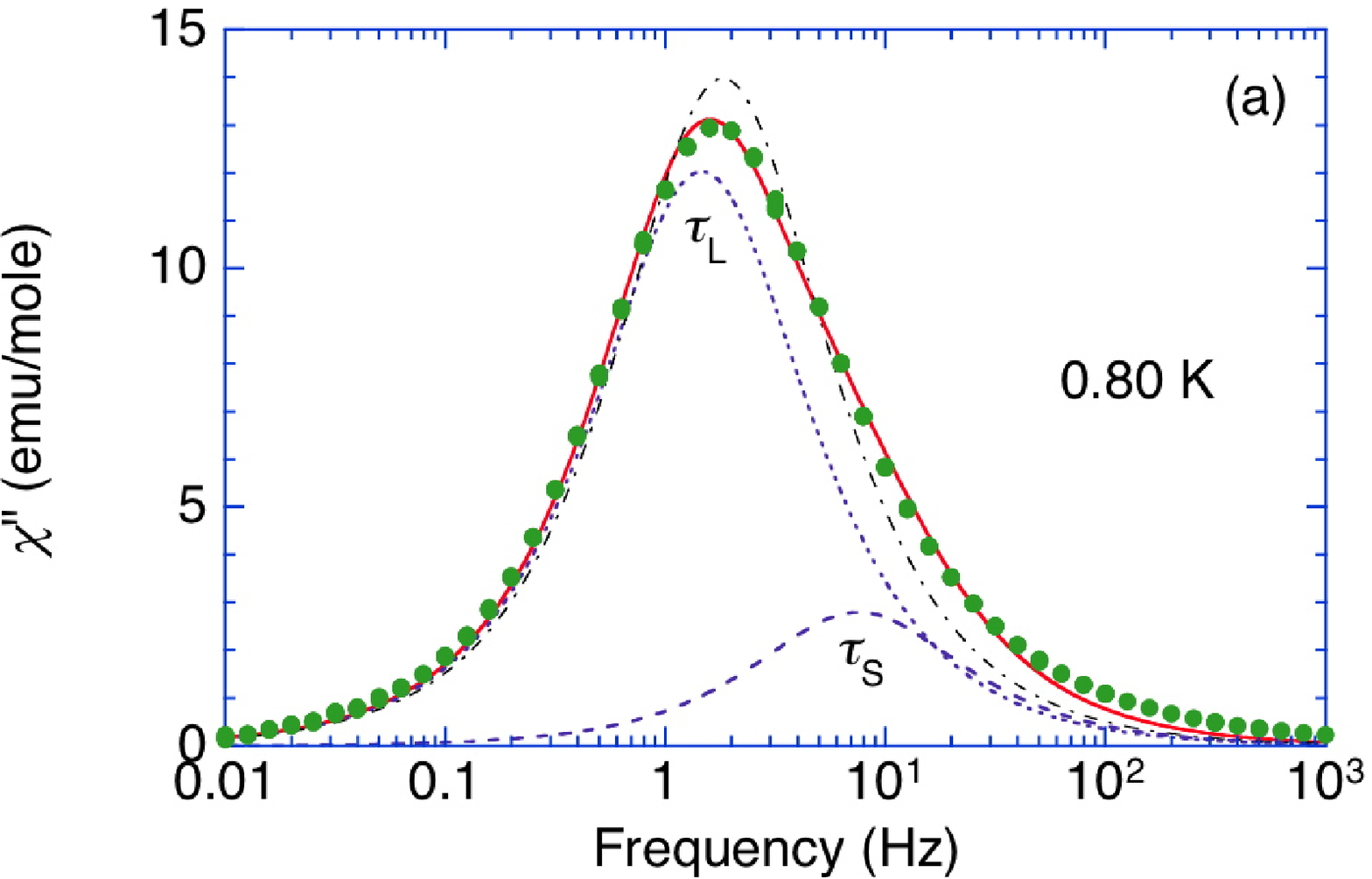}\\
\includegraphics[width=7cm]{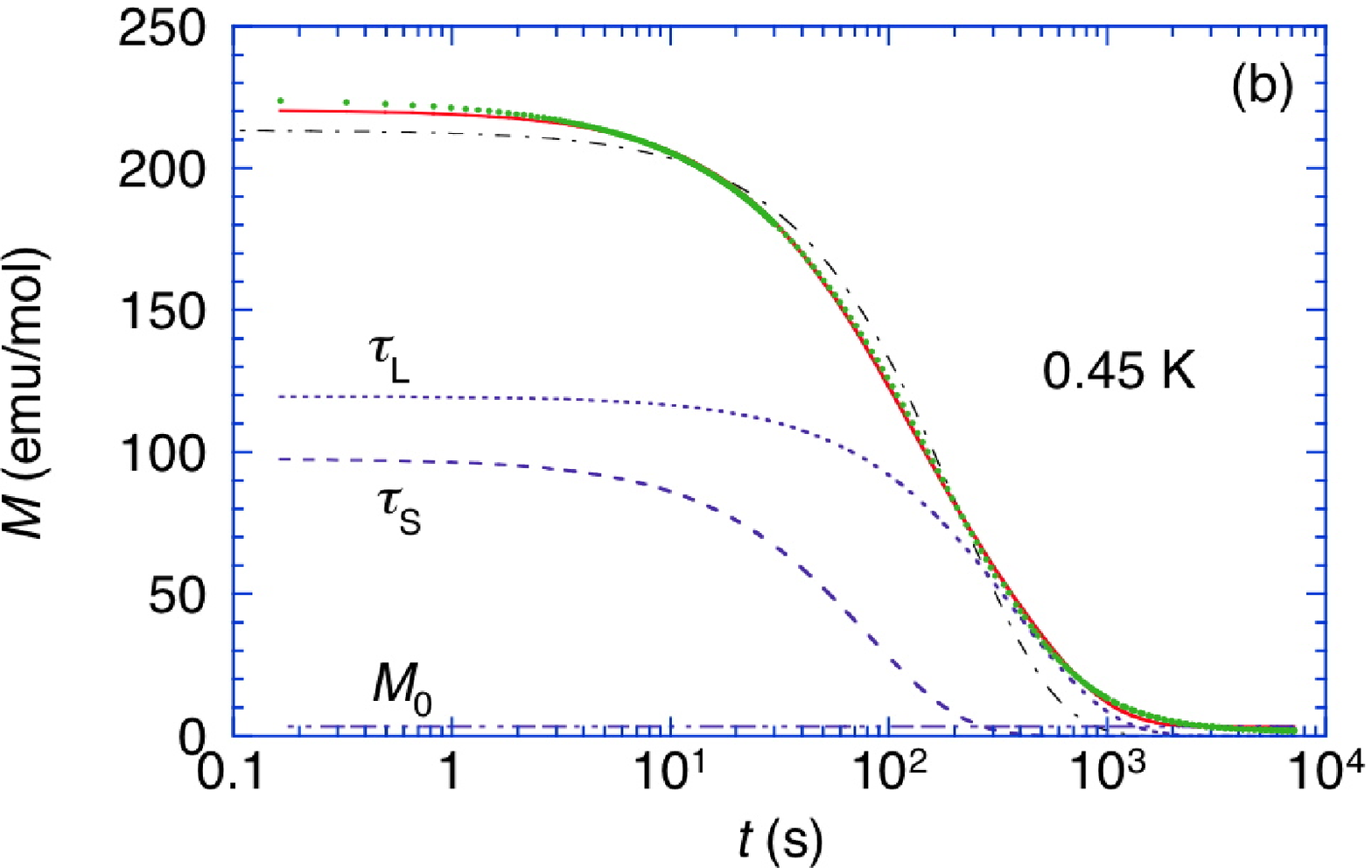}
\end{center}
\caption{
(Color online) Fitting results for (a) $\chi^{\prime\prime}(f)$ at 0.80 K and (b) $M(t)$ at 0.45 K. Solid lines (red) show the fitting curves 
with ($\chi_{\rm L}$, $\chi_{\rm S}$, $\tau_{\rm L}$, $\tau_{\rm S}$) = (24.06, 5.594, 0.1088, 0.02110) for $\chi^{\prime\prime}(f)$ and ($M_{\rm L}$, $M_{\rm S}$, $M_0$, $\tau_{\rm L}$, $\tau_{\rm S}$) = (119.6, 97.60, 3.254, 380.0, 79.72) for $M(t)$, respectively.
Broken and dotted lines show the contribution from $\tau_{\rm S}$ and $\tau_{\rm L}$, respectively.
For comparison, dash-dotted lines (black) show the fitting curves by a single $\tau$ model, which can not reproduce the data.
}
\label{f3}
\end{figure}

\begin{figure}[htbp]
\begin{center}
\includegraphics[width=7cm]{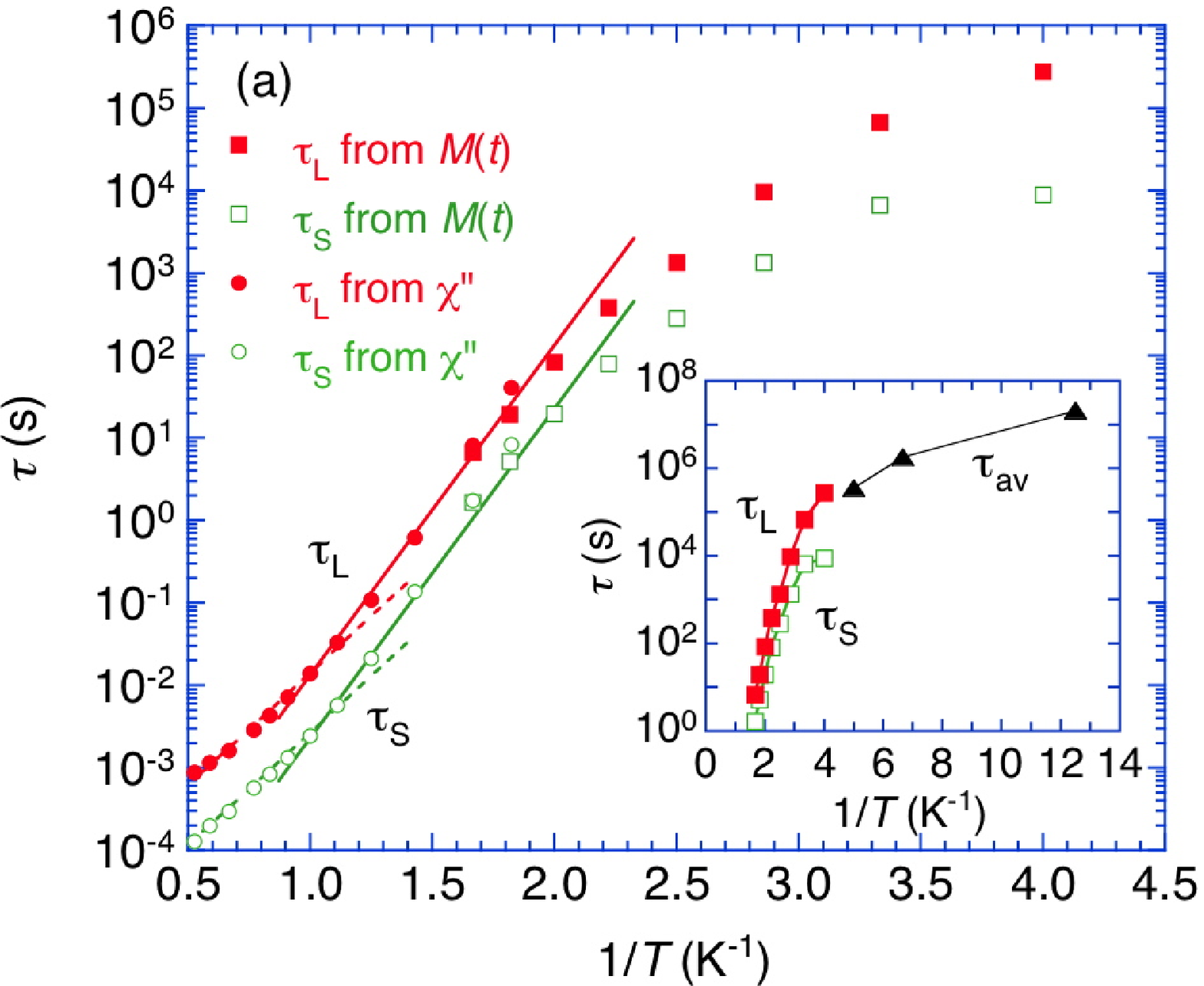}\\
\includegraphics[width=7cm]{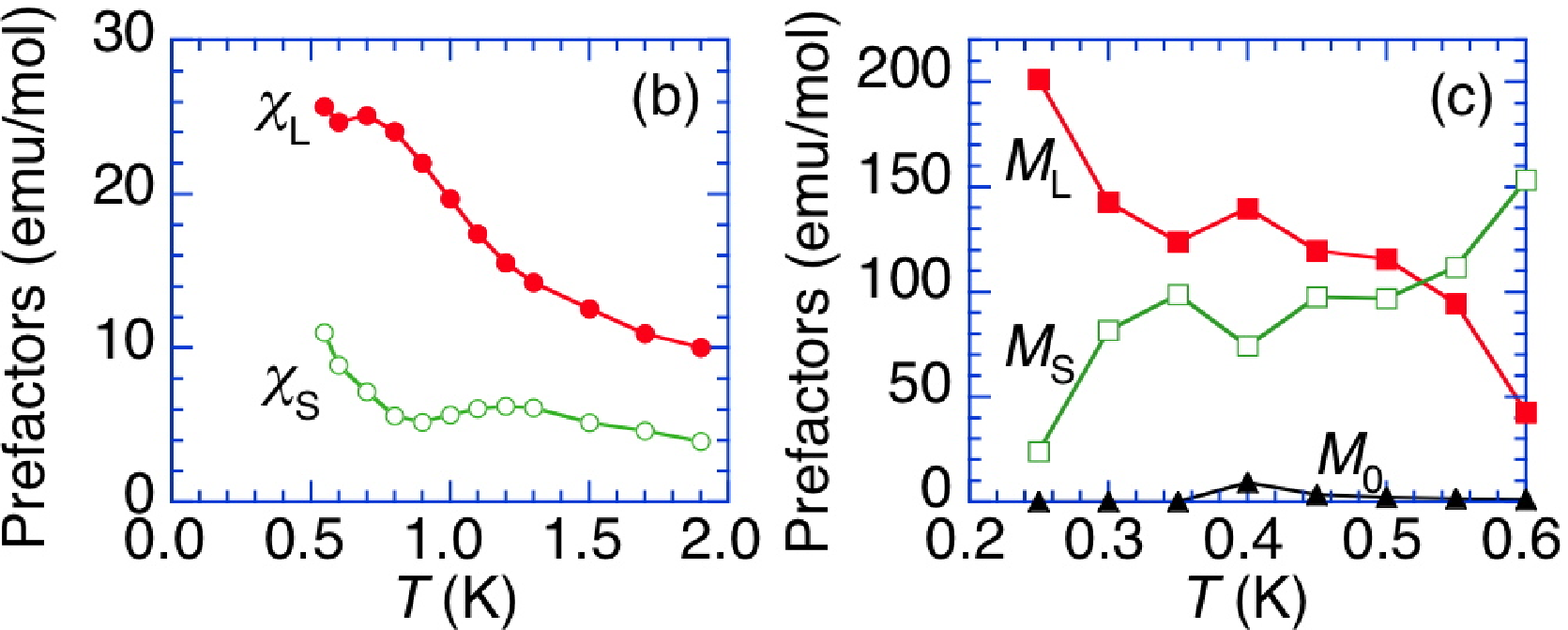}
\end{center}
\caption{
(a)(Color online) Temperature dependence of $\tau_{\rm L}$ and $\tau_{\rm S}$ of Dy$_2$Ti$_2$O$_7$. Solid and broken lines show the Arrhenius law with $E_{\rm p}$ = 9.2 and 6.3 K, respectively. Inset shows the temperature dependence of $\tau_{\rm av}$, $\tau_{\rm L}$ and $\tau_{\rm S}$ down to 0.08 K.(b)(Color online) Temperature dependence of fitting parameters $\chi_{\rm L}$ and $\chi_{\rm S}$. (c)(Color online) Temperature dependence of $M_{\rm L}$, $M_{\rm S}$, and $M_0$. 
}
\label{f4}
\end{figure}

To account for the whole set of measurements,  ($\chi_{\rm AC}$ between 0.50 and 1.9 K and $M(t)$ below 0.60 K), we have fitted the data with a 2-$\tau$ Debye model as follows:
\begin{equation}
\chi^{\prime\prime}(f)=\chi_{\rm L}\frac{2\pi f \tau_{\rm L}}{1+(2\pi f \tau_{\rm L})^2}+\chi_{\rm S}\frac{2\pi f \tau_{\rm S}}{1+(2\pi f \tau_{\rm S})^2},
\label{e1}
\end{equation}
\begin{equation}
M(t)=M_{\rm L} {\rm exp}(-\frac{t}{\tau_{\rm L}})+M_{\rm S} {\rm exp}(-\frac{t}{\tau_{\rm S}})+M_0,
\label{e2}
\end{equation}
where $\chi_{\rm L}$, $\chi_{\rm S}$,  $M_{\rm S}$, $M_{\rm L}$ and $M_0$ are fitting parameters, and $\tau_{\rm L}$ and $\tau_{\rm S}$ are the short and long relaxation times.
Similar shape of $\chi_{\rm AC}$ data in spin ice Ho$_2$Ti$_2$O$_7$ is obtained in the temperature range of 0.5-1.3 K.~\cite{HoTiO-AC}
Figs. \ref{f3}(a) and (b) show the fitting results for $\chi^{\prime\prime}(f)$ at 0.80 K and $M(t)$ at 0.45 K, respectively.
The fits are rather good over a large frequency or time range although they are not perfect and small differences can be seen at higher frequency and at very short or very long times.
However, the fitting results show that the bulk of the relaxation can be reasonably described by two relaxation times.
In addition, the 2-$\tau$ model allows to fit both AC and DC data in a wide temperature range (0.25-1.9 K), and thus to get an estimation of $\tau_{\rm L}(T)$ and $\tau_{\rm S}(T)$ (See Fig. \ref{f4}(a)).
Below 0.25 K, the relaxation of the magnetization is so slow that we cannot fit the data within the 2-$\tau$ model.
However, we can estimate the average relaxation time $\tau_{\rm av}$ using a simple exponential relaxation (See Fig. \ref{f4}(a) inset).
Note that the $\tau_{\rm L}$ and $\tau_{\rm S}$ at 0.55 and 0.60 K obtained from $M(t)$ are in agreement with those from $\chi^{\prime\prime}(f)$, which shows that the 2-$\tau$ model is consistent. 

It is interesting to speculate on the origin of two relaxation times in this system.
To this end, we note that both relaxation times have a similar temperature dependence, at least down to 0.3 K, but their relative contributions to $\chi^{\prime\prime}(f)$ and $M(t)$ depend on temperature. This can be seen from the prefactors $\chi_{\rm L}$, $\chi_{\rm S}$, $M_{\rm L}$ and $M_{\rm S}$ in Figs. \ref{f4}(b) and (c).~\cite{M0}
This observation allows to conclude that the two $\tau$ do not originate from the existence of two different sites in the pyrochlore lattice in [111] applied field.~\cite{KagomeIce}
A recent theoretical study on the basis of magnetic monopoles proposes that two types of magnetic monopoles (free and bound magnetic monopoles) are created when a dipolar spin ice is quenched from high temperature into the spin ice regime.~\cite{ThermalQuench}.
In this model, the free magnetic monopoles have a shorter lifetime as annihilation takes place when two monopoles meet elsewhere, whereas the bound magnetic monopoles have a longer lifetime due to a dynamical arrest.~\cite{ThermalQuench}
However the calculations were made in zero field,~\cite{ThermalQuench} with a very fast cooling rate, presumably much faster than the experiment (10 s to 0.4K, several minutes to 0.1 K), which makes it difficult to compare directly.
Notwithstanding, the two relaxation times in our 2-$\tau$ model may correspond to the lifetime of these free and bound monopoles.

Below 2 K,  $\tau_{\rm L}$ and $\tau_{\rm S}$ have almost the same temperature dependence.
They increase on cooling with the largest rate of change between 0.5 and 1 K.
Surprisingly, the increase of both $\tau_{\rm L}$ and $\tau_{\rm S}$ tends to be suppressed below 0.5 K
but the relaxation times continue to increase although more slowly, at least down to 0.08 K, reaching times as long as 200 days at 0.08 K.

To get an insight for the energy involved in these relaxation process we can describe the $\tau(T)$ dependence with a Arrhenius law:  $\tau(T) = \tau_0 {\rm exp}(E_{\rm p}/T)$ in the temperature range 0.5-1 K.
From the temperature dependence of $\tau_{\rm L}$ and  $\tau_{\rm S}$, we get an energy barrier $E_{\rm p} \approx 9.2$ K ($8 \sim 9J_{\rm eff}$) with $\tau_{0{\rm L}} = 1.35\times 10^{-6}$ s and $\tau_{0 {\rm S}} = 2.3 \times 10^{-7}$ s, which is larger than the previous result $E_{\rm p} \sim 6.6$ K in the temperature range 0.75-2 K.~\cite{DyTiO-AC3, monopole-dynamics}
This Arrhenius fit describes only a narrow temperature range.
A description taking explicitly into account the long range Coulomb interaction between the monopoles is desirable to explain the dynamics below 0.75 K.

A possible origin for  the suppression of the rate of increase of $\tau(T)$ below 0.5 K
could be a reduction of $E_{\rm p}$ at low temperature. The reduction of the barrier comes about because only single defects are  allowed in the low temperature limit.~\cite{monopole-dynamics}
Then, for $\tau_{\rm av}$ below 0.2 K, assuming the Arrhenius law, a rough value of $E_{\rm p} \sim 0.5$ K is obtained.
Obviously, this value is smaller than $2J_{\rm eff}$ expected for a single defect in the nearest neighbor spin ice model.
Therefore, it is hard to describe the dynamics at least below 0.2 K by the Arrhenius law.
As another possible origin is quantum dynamic (QD) effect between two spins.
This QD effect results from the anisotropic superexchange coupling between neighboring Kramers doublets; the QD effect is intrinsic to pyrochlore rare-earth oxides and very significant in Pr pyrochlore oxides.~\cite{TbTiO,TbSnO,PrIrO,Onoda}
The QD of Dirac string derived from the spin-flip exchange interaction may suppress the increasing of $\tau(T)$ at very low temperature.
This effect is expected to be smaller in the Dy$_2$Ti$_2$O$_7$, due to the large spin and the strong anisotropy.
A quantum mechanical treatment of the spins will be necessary to clarifying the dynamics due to QD at very low temperature.
Note that  a similar QD effect is observed in the nuclear quadrupole resonance measurement of Dy$_2$Ti$_2$O$_7$; the  increasing of spin-lattice relaxation time $T_1$ is suppressed below 0.5 K and $T_1$ becomes almost constant below 0.3 K.~\cite{NMR}
These results suggest a QD effect in spin ice.

It is worth noting that our results may be complementary to the recent experiments that report a relaxing magnetic current due to monopoles.~\cite{DyTiO-Giblin}
These relaxation measurements were made after ZFC and then applying a field pulse during 5 to 60 s.
They focused on the short time relaxation ($t< 180$ s).
The latter could be described using a chemical kinetic model, involving dissociation and recombination of the monopoles.
The so-obtained energy barriers at 0.36 K, are 4.5 K for the free monopoles, and 3.3 K for the bound monopoles.
On the contrary, our relaxation curves focus on the way the system reaches equilibrium at long times. 
As seen above, our rough 2-$\tau$ model tends to show that the long time relaxation may also imply free and bound monopoles.
However, fitting the  Arrhenius law, over a very restricted temperature range centered on  0.36 K gives energy barriers of approximately 4 and 4.7 K for the short and long relaxation process respectively, and a very slow $\tau_0$ of approximately $10^{-2}$ s for both.
Although the concentration of each type of monopoles seem to be very different in our experiment, it is not surprising since in our measurements the field was applied at "high" temperature, thus generating different excitations than the ones created at low temperature.

In summary, the $\chi_{\rm AC}$ and $M(t)$ measurements due to field quench after FC along [111] direction have been performed in order to estimate $\tau(T)$ in a spin ice compound Dy$_2$Ti$_2$O$_7$ down to 0.08 K.
We found that the dynamics of spin ice below 2 K is well described by 2-$\tau$ model with $\tau_ {\rm S}$ and $\tau_{\rm L}$.
From the view of magnetic monopole picture, the existence of two relaxation times corresponds to two types of magnetic monopoles (free and bound monopoles).
Below 0.5 K, both $\tau_{\rm S}$ and $\tau_{\rm L}$ show the clear deviation from the thermal activated dynamics with $E_{\rm p} \sim 9$ K. 
The increasing of the relaxation times are suppressed below 0.5 K, suggesting a QD effect.

The authors thank P. Holdsworth, L. D. C. Jaubert, R. Moessner, M. Takigawa and S. Onoda for helpful discussion. 
This work was supported by a Grant-in-Aid for Scientific Research on Priority Areas "Novel States of Matter Induced by Frustration" (No. 19052005 and No. 19052003). 
Crystal growth was performed by using facilities of the Materials Design and Characterization Laboratory in the Institute for Solid State Physics, the University of Tokyo under the Visiting Researcher's Program.


\begin{thebibliography}{99}
\bibitem{SpinIceRev} S. T. Bramwell and M. J. P. Gingras: Science {\bf 294} (2001) 1495.
\bibitem{DyTiO-S} A. P. Ramirez, A. Hayashi, R. J. Cava, R. Siddharthan, and B. S. Shastry: Nature (London) {\bf 399} (1999) 333.
\bibitem{DyTiO-Neutron} T. Fennell, A.A. Petrenko, B. Fak, S. T. Bramwell, M. Enjalran, T. Yavors'kii, M. J. P. Gingras, R. G. Melko and G. Balakrishnan: Phys. Rev. B {\bf 70} (2004) 134408.
\bibitem{SpinIce} M. J. Harris, S. T. Bramwell, D. F. McMorrow, T. Zeiske, and K.W. Godfrey: Phys. Rev. Lett. {\bf 79} (1997) 2554.
\bibitem{Ice} L. Pauling: J. Am. Chem. Soc. {\bf 57} (1935) 2680.
\bibitem{monopole} C. Castelnovo, R. Moessner, and S. L. Sondhi: Nature (London) {\bf 451} (2008) 42.
\bibitem{monopole-Morris} D. J. P. Morris, D. A. Tennant, S. A. Grigera, B. Klemke, C. Castelnovo, R. Moessner, C. Czternasty, M. Meissner, K. C. Rule, J.-U. Hoffmann, K. Kiefer, S. Gerischer, D. Slobinsky and R. S. Perry : Science {\bf 326} (2009) 411.
\bibitem{monopole-Fennell} T. Fennell, P. P. Deen, A. R. Wildes, K. Schmalzl, D. Prabhakaran, A. T. Boothroyd, R. J. Aldus, D. F. McMorrow, and S. T. Bramwell: Science {\bf 326} (2009) 415.
\bibitem{monopole-Kadowaki} H. Kadowaki, N. Doi, Y. Tabata, T. J. Sato, J. W. Lynn, K. Matsuhira and Z. Hiroi: J. Phys. Soc. Jpn. {\bf 78} (2009) 103706.
\bibitem{monopole-Bramwell} S. T. Bramwell, S. R. Giblin, S. Calder, R. Aldus, D. Prabhakaran and T. Fennell: Nature {\bf 461} (2009) 956.
\bibitem{monopole-dynamics} L. D. C. Jaubert and P. C. W. Holdsworth : Nature Phys. {\bf 5} (2009) 258; J Phys.: Condens. Matter {\bf 23} (2011) 164222.
\bibitem{DyTiO-AC} K. Matsuhira, Y. Hinatsu, and T. Sakakibara: J. Phys.: Condens. Matter {\bf 13} (2001) L737.
\bibitem{DyTiO-AC2} J. Snyder, J. S. Slusky, R. J. Cava, and P. Schiffer: Nature {\bf 413} (2001) 48.
\bibitem{DyTiO-AC3} J. Snyder, B. G. Ueland, J. S. Slusky, H. Karunadasa, R. J. Cava, and P. Schiffer: Phys. Rev. B {\bf 69} (2004) 064414.
\bibitem{DyTiO-DC} T. Sakakibra, T. Tayama, Z. Hiroi, K. Matsuhira, and S. Takagi: Phys. Rev. Lett. {\bf 90} (2003) 207205; T. Sakakibara: private communication.
\bibitem{HotSpinIce} G. Ehlers, A. L. Cornelius, M. Orendac, M. Kjnakova, T. Fennell, S. T. Bramwell, and J. S. Gardner: J. Phys. Condens. Matter {\bf 15} (2003) L9.
\bibitem{plateau} The detailed origin of the plateau region is not well-defined. As a possible origin, quantum tunneling process through non-zero off-diagonal components of the dipolar interaction from neighboring spins is considered although thermal assistance is required to make an spin flip.~\cite{HotSpinIce,monopole-dynamics}
\bibitem{DyTiO-MagCalo} M. Orend$\acute{\rm a}\check{\rm c}$, J. Hanko, E. $\check{\rm C}$i$\check{\rm z}$m$\acute{\rm a}$r, A. Orend$\acute{\rm a}\check{\rm c}$ov$\acute{\rm a}$, M. Shirai, and S. T. Bramwell Phys. Rev. B {\bf 75} (2007) 104425.
\bibitem{DyTiO-Giblin} S. R. Giblin, S. T. Bramwell, P. C. W. Holdsworth, D. Prabhakaran and I. Terry: Nature Physics {\bf 7} (2011) 252.
\bibitem{KagomeIce} K. Matsuhira, Z. Hiroi, T. Tayama, S. Takagi, and T. Sakakibara: J. Phys. Condens. Matter {\bf 14} (2002) L559.
\bibitem{Demag} A. Aharoni: J. Appl. Phys. {\bf 83} (1998) 3432.
\bibitem{Previous-data} As there is no description on demagnetization correction in the previous report~\cite{DyTiO-AC3}, the exact comparison is difficult.
\bibitem{DavidsonCole} Davidson-Cole formula can fit the AC data well above 14 K.~\cite{DyTiO-AC}
\bibitem{HoTiO-AC} J. A. Quilliam, L. R. Yaraskavitch, H. A. Dabkowska, B. D. Gaulin, and J. B. Kycia: Phys. Rev. B {\bf 83} (2011) 094424.
\bibitem{M0} The fitting parameter $M_{\rm 0}$ is fixed to zero below 0.35 K to get a better fitting.
\bibitem{ThermalQuench} C. Castelnovo, R. Moessner, and S. L. Sondhi: Phys. Rev. Lett. {\bf 104} (2010) 107201.
\bibitem{TbTiO} J. S. Gardner, B. D. Gaulin, A. J. Berlinsky, P. Waldron, S. R. Dunsiger, N. P. Raju, and J. E. Greedan: Phys. Rev. B {\bf 64} (2001) 224416.
\bibitem{TbSnO} I. Mirebeau, P. Bonville, and M. Hennion: Phys. Rev. B {\bf 76} (2007) 184436.
\bibitem{PrIrO} S. Nakatsuji, Y. Machida, Y. Maeno, T. Tayama, T. Sakakibara, J. van Duijn, L. Balicas, J. N. Millican, R. T. Macaluso and Julia Y. Chan: Phys. Rev. Lett. {\bf 96} (2006) 087204.
\bibitem{Onoda} S. Onoda, and Y. Tanaka: Phys. Rev. Lett. {\bf 105} (2010) 047201; Phys. Rev. B {\bf 83} (2011) 094411.
\bibitem{NMR} K. Kitagawa, K. Ishida, R. Higashinaka, Y. Maeno, and M. Takigawa: Abstr. Int. Conf. Highly Frustrated Magnetism 2008, p. 133.
\end{thebibliography}
\end{document}